\documentclass[letter]{aa}  
\usepackage{graphicx}
\usepackage{txfonts}
\bibpunct{(}{)}{;}{a}{}{,}

\newcommand{\te}{\ensuremath{T_{\rm eff}}}
\newcommand{\bz}{\ensuremath{\langle B_z \rangle}}

\newcommand{\bs}{\ensuremath{\langle \vert B \vert \rangle}}

\newcommand{\wdp}{WD\,0301+059}

\begin{document}

   \title{Discovery of a 
          Sirius-like binary system\\ with a very strongly magnetic white dwarf}

\subtitle{Binarity among magnetic white dwarfs}

   \author{John D. Landstreet
          \inst{1,2}
          \and
          Stefano Bagnulo
          \inst{1}
          }

   \institute{Armagh Observatory and Planetarium, College Hill,
              Armagh BT61 9DG, Northern Ireland, UK \\
              \email{John.Landstreet@Armagh.ac.uk, 
                     Stefano.Bagnulo@Armagh.ac.uk}
         \and
             University of Western Ontario, Department of Physics \& Astronomy,             London, Ontario, Canada N6A 3K7\\
             }

   \date{Received December 11, 2019; accepted January 19, 2020}

 
  \abstract{
Binary systems containing a magnetic white dwarf and a main-sequence star are considered extremely rare, perhaps non-existent. In the course of a search of magnetic fields in high-mass white dwarfs we have discovered a Sirius-like wide binary system composed of a main-sequence G0 star and an $M\sim 1.1\,M_\odot$ white dwarf with a huge (hundreds of MG) magnetic field. This star, WDS J03038+0608B, shows a circular polarisation amplitude of 5\,\% in the continuum, with no evidence of variability in a 1\,d timescale, little or no linear polarisation in the blue part of the spectrum, and about 2\,\% linear polarisation in the red part of the optical spectrum. A search in the literature reveals the existence of at least four more binary systems that include a magnetic white dwarf and a non-degenerate companion; three such systems passed unremarked in previous studies. We estimate  that up to a few percent of magnetic white dwarfs may be found to occur in wide binary pairs. However, at least four of the five known binary systems with a magnetic white dwarf are too widely separated to be expected to evolve into systems experiencing Roche-lobe overflow, and cannot be considered as progenitors of magnetic cataclysmic variable (AM Her and DQ Her) systems.}


   \maketitle
%

\section{Introduction}

Binarity is a fairly common feature of stars in space. The RECONS project\footnote{\tt www.recons.org}, a census of all stars and star systems, brown dwarfs, and planets within 10\,pc of the Sun, estimates that the multiplicity rate (the fraction of star systems containing more than one star) is about 27\,\%. Thus the consequences of membership in multiple stars systems are quite important in the overall understanding of the global effects of stellar evolution.

About 90\,\% of stars end their lives as white dwarfs (WDs). Their subsequent evolution, in most cases, is simply to cool very slowly. The local WD population (e.g. the sample within 20 or
25\,pc of the Sun) contains a fairly unbiassed record of the final products of stellar evolution of all but the most massive stars in this part of the Milky Way galaxy. This sample provides direct
evidence about the rate of production of WDs as a function of time during the roughly 10\,Gyr of star production. It also contains important information about the relative production rates of various specific evolutionary results, such as binary or multiple systems containing one or more WDs, magnetic WDs (MWDs), and WDs polluted by metals due to accretion of debris from former planetary systems \citep{Holbetal16,Holletal18}.

In the samples of nearby WDs, binarity is fairly common. White dwarfs largely occur in three types of binary systems: a WD paired with a low--mass M dwarf (WD+dM); a WD with a higher mass main-sequence (MS) star, usually called a Sirius-like system (Sl); and a system with two WDs (DD). These systems are sometimes resolved on the sky as visual binary systems, but are usually discovered as unresolved binary systems \citep[as composite spectra of two very different stars, e.g. ][]{Lieb09}. Altogether, WD binary and multiple systems account for about 25\,\% of the the total WD population near the Sun \citep{Holbetal13,Toonetal17}. 

Magnetic fields are relatively common in WDs. Recent studies show that about 10\,\% of the current WD population near the Sun reveal the presence of strong ($B > 2$\,MG) magnetic fields \citep[e.g.][]{Kawketal07}, and roughly another 10\,\% show weaker fields between a few kG and 2\,MG \citep{LandBagn19}. However, previous literature has virtually failed to discover MWDs in binary systems in which the WD companion is a MS star. \citet{Liebetal05} and \citet{Liebetal15} have examined the WD--dM composite spectra in the Sloan Digital Sky Survey (SDSS) database, finding no Zeeman splitting in the spectra of the WDs that are members of a close binary system. \citet{Liebetal05} also studied a number of wide common proper motion WD--MS pairs. Although they suspected the presence of magnetic fields in a few of the WD members of such systems, they concluded that the presence of a magnetic field could not be clearly shown in any of the systems they studied. \citet{Liebetal15} confirmed that MWDs had not been found in any WD--MS binary systems, regardless of separation.  The sole clear exception to this situation is found among the relatively rare cataclysmic variables (CVs), close binary systems in which, following an episode of mass transfer, common-envelope evolution, and collapse of one member of the system to a WD, the remaining MS member (usually an M dwarf) actively loses mass onto the WD. Among CVs, about one quarter of WDs have magnetic fields in the MG range \citep{Liebetal15}.  It has repeatedly been pointed out by Liebert and collaborators \citep{Liebetal05,Liebetal15} that this peculiar situation is very difficult to understand. If a substantial fraction of the WDs in CVs have strong magnetic fields that are inherited from their earlier evolution, we would also expect to find a substantial number of MWDs among the WDs in CV precursor and other types of binary systems.

In this paper we report the discovery of a very strong-field MWD that is a member of a visual binary system containing an MS G star. Although this system does not constitute a precursor system to the magnetic CV systems, the general conclusion of \citet{Liebetal15} about the total absence of MWDs in WD--MS binary systems requires qualification.
In addition, we show that a small number of other (visual) binary systems containing an MWD and a
non-degenerate companion have gone largely unremarked in the literature. We conclude by characterising the new, somewhat more complicated, relationship between MWDs and WD binarity.

\section{Observations}\label{Sect_Observations}
It has frequently been argued that MWDs are on average significantly more massive than the (very well-defined) overall WD average mass of $0.6 M_\odot$ \citep{Liebetal03,Ferretal15}. As part of our recent spectropolarimetric studies of WDs \citep[e.g.][]{BagnLand18,LandBagn19}, we have been searching for magnetic fields in high-mass WDs, selecting our candidates with the help of the large database of WDs extracted from the Gaia Data Release 2 (DR2) by \citet{Gentetal19}, and limiting ourselves to stars within roughly 30\,pc of the Sun.
One of our targets was the high proper motion star SDSS J030350.63+060748.9  =  Gaia DR2 6963383233077632. Since the available names for this WD are all long enough to be awkward to use, we refer to this star by the conveniently short but unofficial Villanova name \wdp.

We observed our target using the ISIS spectropolarimeter on the William Herschel 4m telescope. The spectropolarimeter was set up to measure the intensity and polarisation spectrum in both red and blue arms simultaneously, using the configuration adopted by \citet{BagnLand18}.

\begin{table}[]
\caption{Log of spectropolarimetric observations of \wdp}
    \centering
    \begin{tabular}{l l l l l l}
    \hline\hline
    Date        &  UT   &   MJD     & $t_{\rm exp}$ & Stokes & S/N \\
    (yr-mo-day) &       &           &  (s)          & param  & (\AA$^{-1}$) \\
    \hline
    2019-10-08  & 02:32 & 58763.106 & 4800          & $V$    & 200 \\  
    2019-10-09  & 01:50 & 58764.076 & 1800          & $V$    & 100  \\
    2019-10-09  & 02:25 & 58764.101 & 1800          &$Q,U$   &  70  \\
         \hline
    \end{tabular}
    \label{table_2}
\end{table}

Our target is separated by only 12\arcsec\ from a companion that is 8 magnitudes brighter. In spite of the bright light halo, we were able to obtain circular polarisation spectra on two successive nights, and linear polarimetry on one night (see Table~\ref{table_2}). The data were reduced as described by \citet{BagnLand18} for circular polarisation and by \citet{BagnLand19a} for linear polarisation.

\section{Properties of \wdp\ and of its MS companion HD\,19019}
Data about the two stars of this visual binary are tabulated in Table~\ref{table_1}. In the following we discuss binarity and atmospheric characteristics of the two stars.

\begin{table}
\caption{Properties of HD\,19019 -- \wdp\ system members} 
\label{table_1}      
\centering                          
\begin{tabular}{l l l }        
\hline\hline                
Star                   &  HD\,19019        &  \wdp  \\   
Sp                     &   G0V             &  DXP  \\
V                      & 6.93              &    \\
G                      & 6.775             & 14.96  \\
$\alpha$ (ICRS J2000)  & 03 03 50.815      & 03 03 50.56    \\
$\delta$ (ICRS J2000)  & +06 07 59.88      & +06 07 48.75   \\
$\pi$ (mas)            & $31.88 \pm 0.07$  & $31.60 \pm 0.08$  \\
$\mu_{\alpha}$ (mas/yr) & $232.05 \pm 0.14$ & $227.67 \pm 0.14$ \\
$\mu_{\delta}$ (mas/yr) & $49.80\pm 0.15$   & $53.94 \pm 0.13$ \\
\te\,(K)               & $6113 \pm 40$     & $18200 \pm 3000$  \\
$\log g$ (cgs)         & $4.40 \pm 0.1$    & $8.85 \pm 0.15$ \\
Mass ($M_\odot$)       & $1.06 \pm 0.06$   & $1.12 \pm 0.15$  \\
Age (Gyr)              & $3.0 \pm 2.3$     & $0.45 \pm 0.05$  \\
\hline                        
\end{tabular}
\end{table}

\subsection{Binarity}
This very nearby probable WD has quite high proper motion of more than 200\,mas/yr. Remarkably, it is only 12\arcsec\ from a bright MS star, HD\,19019, with very similar values of parallax and both components of proper motion. Formally, the Gaia parallax and proper motions characterising the MS star and the WD are not equal within their stated uncertainties, but the three large and nearly equal pairs of values are all extremely unusual for any random star in the Gaia DR2 catalogue, leading to the conclusion that these two stars do in fact form a physical binary system. The perturbing effect of light from the brighter star has probably slightly affected Gaia astrometry of the secondary, leading to the very small differences in astrometric data reported for the two stars.  
 
The binary nature of this system was apparently first reported by Wilfried Knapp, who has a long series of angular separation and position angle measurements in the Journal of Double Star
Observations, starting with \citet{Knap15}. Knapp has not reported any significant change in the relative coordinates of the two stars in the past four years. The system is  listed in the Washington Double Star Catalogue (as WDS~J03038+0608B).
Using the parallax and angular separation of the two components of the visual
binary, we conclude that the current projected physical separation
between the primary and secondary star is about 380\,AU. 

\subsection{Physical parameters of the stars}
Basic physical parameters of the primary G0V star have been determined by \citet{Neto17} and by \citet{Luck17}. The two studies give similar values for \te\ and $\log g$.  We tabulate Luck's values, and use the differences between the two studies to estimate uncertainties. In
addition, Luck has determined the mass and age of HD\,19019 by comparison with several sets of isochrones. We tabulate his mean value for each, and use the spread in values to estimate uncertainties. The various isochrones give very similar masses, but
because there are small differences in position of the zero-age main sequence (ZAMS) in the Hertzsprung-Russell diagram, the age is not accurately determined. The youngest age reported by Luck for the primary star is 1.4\,Gyr. However, a still younger age of 0.70\,Gyr was obtained by \citet{Casaetal11}. This young age is roughly consistent with the level of chromospheric activity ($\log R'_{\rm HK} \approx -4.67 \pm 0.1$) reported by \citet{Pace13}. Thus there is some (but not strong) evidence that the primary star of the system may be $\la 1$\,Gyr old.   

The physical parameters of the secondary WD are provided by \citet{Gentetal19}, with values computed on two different assumptions: an atmosphere of pure H, or an atmosphere of pure He. As we show below, the line spectrum of \wdp\ does not resemble those of normal WDs of either of these simple surface compositions. Thus neither calibration is expected to provide very accurate physical parameter values for \wdp. The parameter values for the two assumptions differ quite significantly: the H-rich calibration leads to values of $\te = 20823$\,K, $\log g = 8.96$, and $M = 1.18 M_\odot$, while the He-rich calibration leads to $\te = 15545$\,K, $\log g = 8.73$, and $M = 1.05 M_\odot$. Given these large differences, we tentatively assign average values from the two calibrations, with large uncertainties. In any case, it seems clear that \wdp\ is a relatively hot young WD with unusually high gravity and mass. 

Applying the initial-final mass calibration of \citet{Cummetal18}, we find that the parameter values for an H-rich atmosphere suggest an initial ZAMS mass of the progenitor of the WD of about $7.5 M_\odot$, with a lifetime before collapse of roughly $5\,10^7$\,yr  and a cooling age since collapse of about $4\,10^8$\,yr. Assuming the He-rich parameters, the initial mass of the WD was about $6\,M_\odot$, its lifetime as a non-degenerate star was about 1\,$10^8$\,yr, and its cooling age since collapse is about 5\,$10^8$\,yr \citep{Tremetal11,Bergetal11}\footnote{see http://www.astro.umontreal.ca/~bergeron/CoolingModels}. Thus both sets of basic parameters for the WD clearly indicate that if the MWD has evolved as a single star, it started life as an early-B~star with a mass somewhere around 6--8\,$M_\odot$, that the total age of the WD is of the order of 0.5\,Gyr, and therefore that the age of the binary system is also of the order of 0.5\,Gyr. 

The estimated total age of the MWD is somewhat younger than any of the estimates of the age of the primary star. This suggests that the evolution of the WD might have been more complex than effectively single star evolution, for example that the WD may have originated in a close binary system that completed evolution by merging as a single massive WD.

   \begin{figure}
   \centering
\includegraphics[width=9cm,trim={0.4cm 1.5cm 1.32cm 1.8cm},clip]{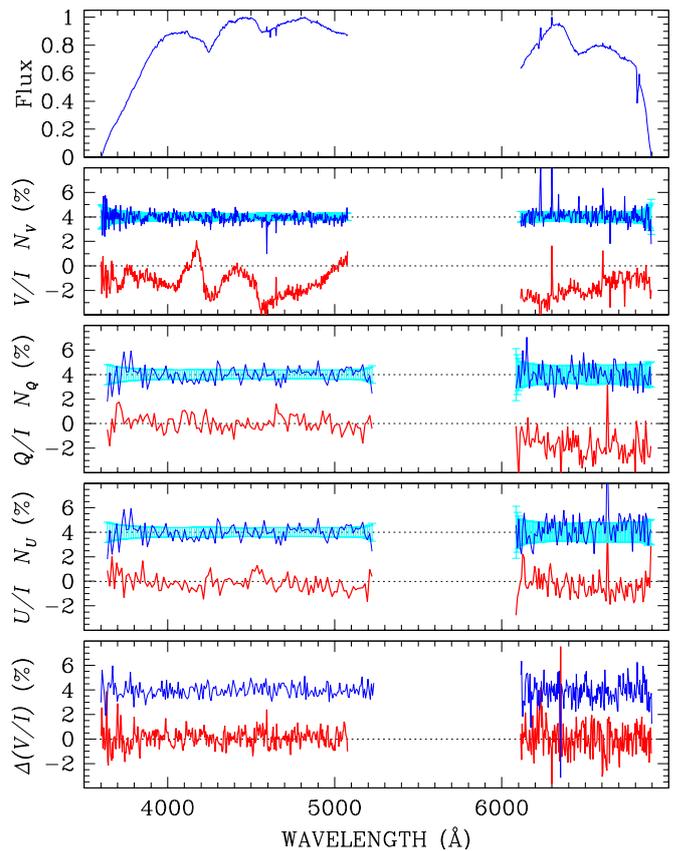}
      \caption{{\it Upper panel:} Observed flux, each segment normalised separately to 1.0, and not corrected for transmission functions. {\it Second panel from top:} $V/I$ spectra obtained on 2019 October 7, and null spectra offset by $+4$\,\% for display purpose. Superposed to the null spectra are the error bars on Stokes $V/I$. {\it Third and fourth panel from the top:} Same as the second panel, but for $Q/I$ and $U/I,$ respectively; data obtained on 2019 October 8. {\it Bottom panel:} Difference between circular polarised spectra obtained on 2019 October 8 and 7, and, offset by $+4$\,\%, the null spectra obtained on 2019 October 7. 
              }
         \label{Fig_circ_poln}
   \end{figure}
\section{Polarised spectrum and magnetic field of \wdp}
The circular polarisation spectrum of \wdp\ from our first night of observation shows the two main hallmarks of a very high-field MWD: the intensity spectrum has asymmetric absorption features at unexpected wavelengths, each 150\,\AA\ or more in full width, and the circular polarisation is non-zero throughout the spectrum, with amplitude varying between about +1.5\% and $-3.5$\%. 
{\em \textup{A priori}}, it was immediately clear that the field of this star probably exceeds 100\,MG. Consequently, we re-observed the circular polarisation on the following night, and obtained a linear polarisation spectrum as well. Polarised spectra of \wdp\ are shown in Fig.~\ref{Fig_circ_poln}.

When we compare the scaled intensity and circular polarisation spectra from the two observations, taken about 23\,hours apart, no significant difference between the two data sets is visible. We have no evidence that the polarisation of this WD varies on a timescale of about 1\,d. 

In the blue spectral window observed, the circular polarisation shows two strong absorption features, one that is deepest at about 4245\,\AA\  and extends from about 4120 to 4360\,\AA\ (and is thus shaded towards shorter wavelengths), and a second feature that is deepest at about 4570\,\AA, extending from about 4525\,\AA\ to about 4800\,\AA, strongly shaded towards longer wavelengths. In the red spectral window, the single flux feature is a broad, shallow  absorption trough centred at about 6530\,\AA, extending between about 6420 and 6610\,\AA, and thus slightly shaded towards shorter wavelengths. In the blue spectrum, each absorption feature corresponds to a strong feature in the circular polarisation spectrum. In the red window, only mild polarisation variations are seen, not particularly centred on the broad absorption trough. 

In linear polarisation, the blue window shows no detectable polarisation signal. However, in the red window, a weak signal of linear polarisation is present throughout the spectrum, slowly increasing from about $p = 2.5$\,\% at 6200\,\AA\ to about $3.0$\,\% amplitude at 6800\,\AA, at a roughly constant position angle of $\Theta \approx 95^\circ$. Since the polarisation is very small or zero in the blue part of the spectrum, in Fig.~\ref{Fig_circ_poln} we have preferred to plot $Q/I$ and $U/I$ rather than linear polarisation and its position angle (which in the blue would be essentially undefined).

As noted above, the presence of several percent of circular polarisation throughout the observed spectrum implies the presence of a magnetic field of the order of 100\,MG or more. In other such high-field WDs, linear polarisation at roughly the 1\,\% level is generally detected as well if the field exceeds perhaps 2--300\,MG. The linear polarisation detected in \wdp\ suggests that the field may be of this magnitude. 

We attempt to identify the dominant chemical element in the atmosphere, and determine the approximate strength of the magnetic field by comparing the wavelengths and widths of the observed flux absorption bands with the positions of the strongest lines computed from atomic physics for stationary transitions, as discussed at length by \citet{Rudeetal94}. 

The dominant spectral features in the available spectral windows are the two strong absorption features extending over 4120--4369\,\AA\ and 4525--4800\,\AA, both of which correspond to strong  local variations in the circular polarisation. Similar features are visible in the flux spectra of a number of very high-field MWDs, including Grw+70$^\circ$\,8247 = WD\,1900+705 \citep{BagnLand19a}, PG\,1031+234 = WD\,1031+234 \citep{Schmetal86}, SBSS\,1349+545 = WD\,1345+545 \citep{Liebetal94}, and SDSS\,1206+0813 = WD\,1203+085 \citep{Vanletal05}. The positions of these two strong features vary  by tens of \AA\ from one star to another. In general, these two features are attributed to two remarkably stable (stationary) components of H$\beta$, due to transitions that at zero field are described as 4f'$_{-1} \rightarrow 2$s$_0$ and 4f'$_{0} \rightarrow 2$s$_0$. For fields lower than about 1000\,MG, the wavelengths of these two transitions shift back and forth by only a few hundred \AA, which means that even substantial variations of field strength over the visible stellar surface (such as those present in a dipolar field distribution) do not smear absorption lines produced locally by these transitions over a very wide interval. As a result, these transitions are robust indicators of a hydrogen atmosphere, and by comparing the observed doublet feature to the computed line positions, it is possible to estimate a range of magnetic field strength present on the visible stellar surface \citep{Rudeetal94}. 

Similarly, two transitions of the H$\alpha$ magnetically split components due to 3p$_{0} \rightarrow 2$s$_0$ and 3d$_{-2} \rightarrow 2$p$_{-1}$ remain within about 1000\,\AA\ of their zero-field wavelengths up to fields of the order of 1000\,MG. They can sometime also be used as indicators of both composition and approximate field strength.  

Without the tools to compute the huge model grids that are needed to search for plausible fits to the data we have acquired, we can at best make qualitative arguments for the range of magnetic field strengths on the surface of \wdp. It appears that a field with strength ranging between about 200 and 250\,MG in an H-rich atmosphere would be expected to introduce strong line absorption from stationary H$\beta$ components in the wavelength range where the two absorption bands are observed \citep{Rudeetal94}. Furthermore, a field strength of this magnitude would probably be consistent with the linear polarisation we detect. Fields of this magnitude or larger seem usually to be associated with detectable levels of linear polarisation, while fields of the order of 100\,MG or lower generally lack detectable linear polarisation \citep[e.g.][Table\,2]{Land92}. 

The main weakness of our suggested range of field strengths is that it does not offer an obvious explanation for the shallow absorption trough at 6500\,\AA. However, we have not found any range of field strengths that provides a more satisfactory global explanation for our data. 

\section{Binary systems with an MS star and an MWD}
\begin{table*}[t]
\caption{MWDs in binary star systems} 
\label{table_3}      
\centering                          
\begin{tabular}{l l r r l r r c c l l}        
\hline\hline               
Villanova & Spectral & Mass & $V$ & dist &$\langle \vert B \vert \rangle$&$\langle B_z\rangle$& System & MS & proj sep & References  \\ 
WD name   & class & ($M_\odot$) & (mag) & (pc) & (MG)    & (MG)   & type   & star & (AU) &   \\
\hline                   
0301+059 & DA(?)P   & 1.1   & 15    & 31.4 & 200?      &        &  Sirius-like    &  G0V & 377   &this work  \\
1009-184 & DZH8.5 & 0.53    & 15.44 & 18.1 & $<0.3$  & ~0.05  &  Sirius-like    &  K7  & 7240 & H13, H16, T17, B19 \\
1105-048 & DAH3.5 & 0.60    & 12.92 & 24.8 & $<0.02$ & ~0.003 &  WD+dM      &  M3  & 6920 & A04, H16, T17, B18  \\
2154-512 & DQH8.3 & 0.60    & 14.74 & 14.9 & $>4$    &  1.3   &  WD+dM    &  M2  & 417  & V10, H16, T17  \\
NGC\,2422 & DBH   & 1.06    &19.8 (G) & 500 & 2.5 &       &  WD+dM    &  dM: & $<250$ & R19 \\
\hline                     
\end{tabular}
\tablefoot{ 
Key to references: K01: \citet{Koesetal01}; A04: \citet{Aznaetal04}; E06: Eisenstein et al 2006;
K09: \citet{Koesetal09}; V10: \citet{Vornetal10}; H13: \citet{Holbetal13}; H16: \citet{Holbetal16};
T17: \citet{Toonetal17}; B18: \citet{BagnLand18}; B19: \citet{BagnLand19b}; R19: \citet{Richetal19}. }
\end{table*}

We have already shown that the broad statement by \citet{Liebetal15} about the complete absence of MWDs in binary systems together with MS companions is inconsistent with the system that we have discussed above, but the more restrictive form of this assertion, that there are no known close binary precursors to the CV systems that contain MWDs, still seems to be valid. We next consider what recent discoveries of MWDs have revealed about other binary systems combining an MS star and an MWD. In fact, in the literature we found another four of these systems, although only one of these was explicitly recognised and discussed. All these systems are listed in Table~\ref{table_3}, which gives the Villanova WD designation (if available); the spectral class, the mass and the $V$ (or $G$) magnitude of the WD; the distance of the system; an estimate of the typical surface magnetic field strength as measured by the mean field modulus (if possible); a typical value of the mean longitudinal field \bz\ (if measured); the system type (Sirius-like if the MS primary is spectral type K or earlier, WD+dM otherwise); the spectral type of the non-degenerate companion; the projected linear separation of the two stars in AU; and references as noted in the footnote to the table. Three of these systems are included in the catalogue of \citet{Toonetal17} of nearby binary systems that contain a WD and an MS star. In the case of WD\,2154--512, \citet{Toonetal17} indicate in their Table~1 that the system contains an MWD (spectral class DQP8.3) and include a citation to field measurement by \citet{Vornetal10}, without explicitly noting that this is one of the rare MWD--dM binaries. One other MS -- WD binary in the same Table~1 of \citet{Toonetal17}, WD\,1105-048, has been known for some years to contain an MWD, although this WD has what is probably the weakest WD field so far detected \citep{Aznaetal04,BagnLand18}. A third entry in the table, WD\,1009-184, includes a very weak-field DZ only recently discovered as magnetic by \citet{BagnLand19b}.

A final entry in Table~\ref{table_3} is the DBA MWD--dM binary recently discovered in the open cluster NGC\,2422 by \citet{Richetal19}. This system is explicitly discussed by the authors as a possible counter-example to the assertion by \citet{Liebetal15}, although at a distance of 500\,pc, it is not completely certain that the system is actually a physical binary, and we will not know if the NGC\,2422 binary could be a precursor to a CV system until a good series of radial velocity measurements of the WD is available. It is interesting to note that except for the different atmospheric chemistry, the system WD\,2154--512, placed at 500\,pc distance, would look very similar to the system in NGC\,2422: the binary system consists of an MWD with a dM companion; the MWD has a field of a few MG; and with only about 1\arcsec\ separation, WD\,2154--512 would be almost unresolved on the sky. It is possible that the NGC\,2422 binary is similar to the local widely separated MWD -- MS star systems.

Thus it is clear that although only a few binary systems including an MWD and a non-degenerate star have been identified, several are found even within the solar neighbourhood. With two such systems within the 20\,pc volume around the Sun, which contains a total of nearly 30 MWDs, as much as a few percent of all MWDs may reside in wide MWD -- MS binary systems. None of the known MWD -- MS binary systems, with the possible exception of the system in NGC\,2422, will become CVs in the future. The contention of \citet{Liebetal15} about the total non-occurrence of MWDs in binary systems should now be restricted to the close binary systems that may be precursors of CVs.

Does our study shed a new light on the theory \citep{Touetal08,Brietal15} that magnetic fields of WDs are generated during the merging of a binary system? We found in binary stars a situation that in fact is very similar to that found among isolated field MWDs, some of which have a higher than average mass, but some have not. Two of the binary systems of Table~\ref{table_3} have a massive MWD that could have been formed as a result of a merger in triple systems that were originally a close binary pair and a distant companion; the remaining three systems have WDs of average mass. 

\section{Conclusions}

In the course of a spectropolarimetric search for magnetic fields in particularly massive WDs, we found continuum circular polarisation, indicating a field of the order of 100\,MG or more, in \wdp, a degenerate star close to a G0V MS star on the sky. Because the measured parallax and proper motion components of the two stars are unusual and very similar, we consider that the WD and the G star nearby form a physical Sirius-like binary system, although the stars in the system are widely separated.  This type of binary system, with an MWD paired with an MS star, is extremely rare. 

The circular and linear polarisation are similar to values found in other MWDs showing continuum polarisation, with circular polarisation values of 2 or 3\,\%.
\wdp\ also shows linear polarisation in the spectral region around H$\alpha$, but no detectable linear polarisation in the blue. 
Because of the presence of significant levels of circular and linear polarisation, we have tried to estimate the global magnetic field strength of \wdp\ by comparing the absorption features in the flux spectrum with the wavelengths of stationary spectral lines as predicted by calculations of atomic levels in large fields \citep{Rudeetal94}. We believe that the data favour a mean field strength \bs\ of roughly 200 -- 250\,MG. 

Based on the rather common occurrence of MWDs in cataclysmic variable (close binary systems containing a WD accreting gas from a very nearby MS companion, usually an M star),  \citet{Liebetal03,Liebetal05,Liebetal15} have carried out very large-scale searches for the progenitors of such systems, searching for binary systems containing an MWD and an MS separated widely enough to be non interacting at present. These searches have been uniformly unsuccessful, and a few years ago,  \citet{Liebetal15} reported that as far as they could determine, there were no MWDs paired with non-degenerate companions.

Very recently, \citet{Richetal19} reported the probable discovery of such a system in the open cluster NGC\,2422. We have shown that three further wide binary systems, each containing an MWD and an MS star, have already appeared in the literature without having been noticed as examples of this binary system type. We are aware of at least five binary systems containing an MWD and an MS star (see Table\,\ref{table_3}).  Two of these systems are within the 20\,pc volume around the Sun. Since this volume contains roughly 30 MWDs, it appears that as many as a few percent of MWDs may appear in such wide binary pairs. However, all five MWD - MS binary systems known are (or in the case of the NGC\,2422 WD, may well be) far too widely separated to be expected to evolve into systems experiencing Roche-lobe overflow, and are thus not{\em } the missing progenitors of CV systems sought by \citet{Liebetal15}.

\begin{acknowledgements}
This work is based on observations collected at the William Herschel
Telescope, operated on the island of La Palma by the Isaac Newton
Group, programme P8 during semester 19B. JDL acknowledges the financial
support of the Natural Sciences and Engineering Research Council of Canada
(NSERC), funding reference number 6377-2016. This research has made use of the Washington Double Star Catalog maintained at the U.S. Naval Observatory. 
\end{acknowledgements}
\bibliography{sirius}

\begin{thebibliography}{35}
\expandafter\ifx\csname natexlab\endcsname\relax\def\natexlab#1{#1}\fi

\bibitem[{{Aznar Cuadrado} {et~al.}(2004){Aznar Cuadrado}, {Jordan},
  {Napiwotzki}, {Schmid}, {Solanki}, \& {Mathys}}]{Aznaetal04}
{Aznar Cuadrado}, R., {Jordan}, S., {Napiwotzki}, R., {et~al.} 2004, \aap, 423,
  1081

\bibitem[{{Bagnulo} \& {Landstreet}(2018)}]{BagnLand18}
{Bagnulo}, S. \& {Landstreet}, J.~D. 2018, \aap, 618, A113

\bibitem[{{Bagnulo} \& {Landstreet}(2019{\natexlab{a}})}]{BagnLand19b}
{Bagnulo}, S. \& {Landstreet}, J.~D. 2019{\natexlab{a}}, \aap, 630, A65

\bibitem[{{Bagnulo} \& {Landstreet}(2019{\natexlab{b}})}]{BagnLand19a}
{Bagnulo}, S. \& {Landstreet}, J.~D. 2019{\natexlab{b}}, \mnras, 486, 4655

\bibitem[{{Bergeron} {et~al.}(2011){Bergeron}, {Wesemael}, {Dufour},
  {Beauchamp}, {Hunter}, {Saffer}, {Gianninas}, {Ruiz}, {Limoges}, {Dufour},
  {Fontaine}, \& {Liebert}}]{Bergetal11}
{Bergeron}, P., {Wesemael}, F., {Dufour}, P., {et~al.} 2011, \apj, 737, 28

\bibitem[{{Briggs} {et~al.}(2015){Briggs}, {Ferrario}, {Tout},
  {Wickramasinghe}, \& {Hurley}}]{Brietal15}
{Briggs}, G.~P., {Ferrario}, L., {Tout}, C.~A., {Wickramasinghe}, D.~T., \&
  {Hurley}, J.~R. 2015, \mnras, 447, 1713

\bibitem[{{Casagrande} {et~al.}(2011){Casagrande}, {Sch{\"o}nrich}, {Asplund},
  {Cassisi}, {Ram{\'\i}rez}, {Mel{\'e}ndez}, {Bensby}, \&
  {Feltzing}}]{Casaetal11}
{Casagrande}, L., {Sch{\"o}nrich}, R., {Asplund}, M., {et~al.} 2011, \aap, 530,
  A138

\bibitem[{{Cummings} {et~al.}(2018){Cummings}, {Kalirai}, {Tremblay},
  {Ramirez-Ruiz}, \& {Choi}}]{Cummetal18}
{Cummings}, J.~D., {Kalirai}, J.~S., {Tremblay}, P.~E., {Ramirez-Ruiz}, E., \&
  {Choi}, J. 2018, \apj, 866, 21

\bibitem[{{Ferrario} {et~al.}(2015){Ferrario}, {de Martino}, \&
  {G{\"a}nsicke}}]{Ferretal15}
{Ferrario}, L., {de Martino}, D., \& {G{\"a}nsicke}, B.~T. 2015, \ssr, 191, 111

\bibitem[{{Gentile Fusillo} {et~al.}(2019){Gentile Fusillo}, {Tremblay},
  {G{\"a}nsicke}, {Manser}, {Cunningham}, {Cukanovaite}, {Hollands}, {Marsh},
  {Raddi}, {Jordan}, {Toonen}, {Geier}, {Barstow}, \& {Cummings}}]{Gentetal19}
{Gentile Fusillo}, N.~P., {Tremblay}, P.-E., {G{\"a}nsicke}, B.~T., {et~al.}
  2019, \mnras, 482, 4570

\bibitem[{{Holberg} {et~al.}(2013){Holberg}, {Oswalt}, {Sion}, {Barstow}, \&
  {Burleigh}}]{Holbetal13}
{Holberg}, J.~B., {Oswalt}, T.~D., {Sion}, E.~M., {Barstow}, M.~A., \&
  {Burleigh}, M.~R. 2013, \mnras, 435, 2077

\bibitem[{{Holberg} {et~al.}(2016){Holberg}, {Oswalt}, {Sion}, \&
  {McCook}}]{Holbetal16}
{Holberg}, J.~B., {Oswalt}, T.~D., {Sion}, E.~M., \& {McCook}, G.~P. 2016,
  \mnras, 462, 2295

\bibitem[{{Hollands} {et~al.}(2018){Hollands}, {Tremblay}, {G{\"a}nsicke},
  {Gentile-Fusillo}, \& {Toonen}}]{Holletal18}
{Hollands}, M.~A., {Tremblay}, P.~E., {G{\"a}nsicke}, B.~T., {Gentile-Fusillo},
  N.~P., \& {Toonen}, S. 2018, \mnras, 480, 3942

\bibitem[{{Kawka} {et~al.}(2007){Kawka}, {Vennes}, {Schmidt}, {Wickramasinghe},
  \& {Koch}}]{Kawketal07}
{Kawka}, A., {Vennes}, S., {Schmidt}, G.~D., {Wickramasinghe}, D.~T., \&
  {Koch}, R. 2007, \apj, 654, 499

\bibitem[{{Knapp}(2015)}]{Knap15}
{Knapp}, W. 2015, Journal of Double Star Observations, 11, 384

\bibitem[{{Koester} {et~al.}(2001){Koester}, {Napiwotzki}, {Christlieb},
  {Drechsel}, {Hagen}, {Heber}, {Homeier}, {Karl}, {Leibundgut}, {Moehler},
  {Nelemans}, {Pauli}, {Reimers}, {Renzini}, \& {Yungelson}}]{Koesetal01}
{Koester}, D., {Napiwotzki}, R., {Christlieb}, N., {et~al.} 2001, \aap, 378,
  556

\bibitem[{{Koester} {et~al.}(2009){Koester}, {Voss}, {Napiwotzki},
  {Christlieb}, {Homeier}, {Lisker}, {Reimers}, \& {Heber}}]{Koesetal09}
{Koester}, D., {Voss}, B., {Napiwotzki}, R., {et~al.} 2009, \aap, 505, 441

\bibitem[{{Landstreet}(1992)}]{Land92}
{Landstreet}, J.~D. 1992, \aapr, 4, 35

\bibitem[{{Landstreet} \& {Bagnulo}(2019)}]{LandBagn19}
{Landstreet}, J.~D. \& {Bagnulo}, S. 2019, \aap, 628, A1

\bibitem[{{Liebert}(2009)}]{Lieb09}
{Liebert}, J. 2009, in Journal of Physics Conference Series, Vol. 172, Journal
  of Physics Conference Series, 012040

\bibitem[{{Liebert} {et~al.}(2003){Liebert}, {Bergeron}, \&
  {Holberg}}]{Liebetal03}
{Liebert}, J., {Bergeron}, P., \& {Holberg}, J.~B. 2003, \aj, 125, 348

\bibitem[{{Liebert} {et~al.}(2015){Liebert}, {Ferrario}, {Wickramasinghe}, \&
  {Smith}}]{Liebetal15}
{Liebert}, J., {Ferrario}, L., {Wickramasinghe}, D.~T., \& {Smith}, P.~S. 2015,
  \apj, 804, 93

\bibitem[{{Liebert} {et~al.}(1994){Liebert}, {Schmidt}, {Lesser}, {Stepanian},
  {Lipovetsky}, {Chaffe}, {Foltz}, \& {Bergeron}}]{Liebetal94}
{Liebert}, J., {Schmidt}, G.~D., {Lesser}, M., {et~al.} 1994, \apj, 421, 733

\bibitem[{{Liebert} {et~al.}(2005){Liebert}, {Wickramasinghe}, {Schmidt},
  {Silvestri}, {Hawley}, {Szkody}, {Ferrario}, {Webbink}, {Oswalt}, {Smith}, \&
  {Lemagie}}]{Liebetal05}
{Liebert}, J., {Wickramasinghe}, D.~T., {Schmidt}, G.~D., {et~al.} 2005, \aj,
  129, 2376

\bibitem[{{Luck}(2017)}]{Luck17}
{Luck}, R.~E. 2017, \aj, 153, 21

\bibitem[{{Netopil}(2017)}]{Neto17}
{Netopil}, M. 2017, \mnras, 469, 3042

\bibitem[{{Pace}(2013)}]{Pace13}
{Pace}, G. 2013, \aap, 551, L8

\bibitem[{{Richer} {et~al.}(2019){Richer}, {Kerr}, {Heyl}, {Caiazzo},
  {Cummings}, {Bergeron}, \& {Dufour}}]{Richetal19}
{Richer}, H.~B., {Kerr}, R., {Heyl}, J., {et~al.} 2019, \apj, 880, 75

\bibitem[{{Ruder} {et~al.}(1994){Ruder}, {Wunner}, {Herold}, \&
  {Geyer}}]{Rudeetal94}
{Ruder}, H., {Wunner}, G., {Herold}, H., \& {Geyer}, F. 1994, {Atoms in Strong
  Magnetic Fields. Quantum Mechanical Treatment and Applications in
  Astrophysics and Quantum Chaos}

\bibitem[{{Schmidt} {et~al.}(1986){Schmidt}, {West}, {Liebert}, {Green}, \&
  {Stockman}}]{Schmetal86}
{Schmidt}, G.~D., {West}, S.~C., {Liebert}, J., {Green}, R.~F., \& {Stockman},
  H.~S. 1986, \apj, 309, 218

\bibitem[{{Toonen} {et~al.}(2017){Toonen}, {Hollands}, {G{\"a}nsicke}, \&
  {Boekholt}}]{Toonetal17}
{Toonen}, S., {Hollands}, M., {G{\"a}nsicke}, B.~T., \& {Boekholt}, T. 2017,
  \aap, 602, A16

\bibitem[{{Tout} {et~al.}(2008){Tout}, {Wickramasinghe}, {Liebert}, {Ferrario},
  \& {Pringle}}]{Touetal08}
{Tout}, C.~A., {Wickramasinghe}, D.~T., {Liebert}, J., {Ferrario}, L., \&
  {Pringle}, J.~E. 2008, \mnras, 387, 897

\bibitem[{{Tremblay} {et~al.}(2011){Tremblay}, {Bergeron}, \&
  {Gianninas}}]{Tremetal11}
{Tremblay}, P.~E., {Bergeron}, P., \& {Gianninas}, A. 2011, \apj, 730, 128

\bibitem[{{Vanlandingham} {et~al.}(2005){Vanlandingham}, {Schmidt},
  {Eisenstein}, {Harris}, {Anderson}, {Hall}, {Liebert}, {Schneider},
  {Silvestri}, {Stinson}, \& {Wolfe}}]{Vanletal05}
{Vanlandingham}, K.~M., {Schmidt}, G.~D., {Eisenstein}, D.~J., {et~al.} 2005,
  \aj, 130, 734

\bibitem[{{Vornanen} {et~al.}(2010){Vornanen}, {Berdyugina}, {Berdyugin}, \&
  {Piirola}}]{Vornetal10}
{Vornanen}, T., {Berdyugina}, S.~V., {Berdyugin}, A.~V., \& {Piirola}, V. 2010,
  \apjl, 720, L52

\end{thebibliography}
\end{document}